\documentclass{iau}
\usepackage{graphicx}

\usepackage{txfonts}
\usepackage{color}
\usepackage{colortab}
\usepackage{pstricks}
\usepackage{enumerate}
\usepackage[normalem]{ulem}

\title[Resonant switch model] 
{Resonant switch model of twin peak HF~QPOs applied to the~atoll source 4U~1636$-$53}

\author[Zden\v{e}k Stuchl\'{\i}k, Andrea Kotrlov\'a \& Gabriel T\"{o}r\"{o}k]   
{Zden\v{e}k Stuchl\'{\i}k, Andrea Kotrlov\'a \and Gabriel T\"{o}r\"{o}k}

\affiliation{Institute of Physics, Faculty of Philosophy and Science, Silesian University in Opava,\\
Bezru\v{c}ovo n\'{a}m. 13, CZ-74601 Opava, Czech Republic\\ email: {\tt zdenek.stuchlik@fpf.slu.cz}, {\tt andrea.kotrlova@fpf.slu.cz} \\[\affilskip]
}

\pubyear{2012}
\volume{290}  
\jname{Feeding compact objects: Accretion on all scales}
\editors{C.M. Zhang, T. Belloni, M. M\'endez \& S.N. Zhang, eds.}

\begin{document}

\maketitle

\begin{abstract}
We present a Resonant Switch (RS) model of twin peak high-frequency quasi-periodic oscillations (HF~QPOs), assuming switch of twin oscillations at a~resonant point, where frequencies of the~upper and lower oscillations $\nu_{\mathrm{U}}$ and $\nu_{\mathrm{L}}$ become to be commensurable and the~twin oscillations change from one pair of the~oscillating modes (corresponding to a~specific model of HF~QPOs) to some other pair due to non-linear resonant phenomena. The RS model enables to determine range of allowed values of spin $a$ and mass $M$ of the~neutron star located at the~atoll source 4U~1636$-$53 where two resonant points are observed at frequency ratios $\nu_{\mathrm{U}} : \nu_{\mathrm{L}} = 3\!:\!2$, 5\,:\,4.
\end{abstract}

\firstsection 
\section{Introduction}

In the~case of LMXBs containing neutron or quark stars
we propose a new model of twin peak HF QPOs assuming switching of the twin oscillatory modes creating sequences of the lower and upper HF QPOs at a resonant point. According to such a Resonant Switch (RS) model non-linear resonant phenomena will cause excitation of a new oscillatory mode (or two new oscillatory modes) and vanishing of one of the previously acting modes (or both the previous modes), i.e., switching from one pair of the oscillatory modes to other pair of them that will be acting up to the following resonant point~\cite{Stu-Kot-Tor:2012:AcA}.

\section{Resonant switch model of HF~QPOs in neutron star systems}

We assume two resonant points at the~disc radii $r_{\mathrm{out}}$ and $r_{\mathrm{in}}$, with observed frequencies $\nu_{\mathrm{U}}^{\mathrm{out}}$, $\nu_{\mathrm{L}}^{\mathrm{out}}$ and $\nu_{\mathrm{U}}^{\mathrm{in}}$, $\nu_{\mathrm{L}}^{\mathrm{in}}$, being in commensurable ratios
$p^{\mathrm{out}} = n^{\mathrm{out}}: m^{\mathrm{out}}$ and $p^{\mathrm{in}} = n^{\mathrm{in}}: m^{\mathrm{in}}$. These resonant frequencies are determined by the~energy switch effect (\cite[T{\"{o}}r{\"{o}}k 2009]{Tor:2009:ASTRA:ReversQPOs}); observations put the~restrictions $\nu_{\mathrm{U}}^{\mathrm{in}} > \nu_{\mathrm{U}}^{\mathrm{out}}$ and $p^{\mathrm{in}} < p^{\mathrm{out}}$. In the~region covering the~resonant point at $r_{\mathrm{out}}$ the~twin oscillatory modes with the~upper (lower) frequency are determined by the~function $\nu_{\mathrm{U}}^{\mathrm{out}}(x;M,a)$ ($\nu_{\mathrm{L}}^{\mathrm{out}}(x;M,a)$). Near the~inner resonant point at $r_{\mathrm{in}}$ different oscillatory modes given by the~frequency functions $\nu_{\mathrm{U}}^{\mathrm{in}}(x;M,a)$ and $\nu_{\mathrm{L}}^{\mathrm{in}}(x;M,a)$ occur ($x\equiv r/M$ is the~dimensionless radius).

We assume all the~frequency functions to be determined by combinations of the~orbital and epicyclic frequencies of the~geodesic motion in the~Kerr backgrounds. Such a~simplification is correct with high precision for neutron (quark) stars with large masses, close to the maximum mass allowed for a~given equation of state. The~frequency functions have to meet the~observationally given resonant frequencies. In the~framework of the~simple RS~model, when two resonant points and two pairs of the~frequency functions are assumed, this requirement enables determination of the~neutron (quark) star parameters. The~``shooting'' of the~frequency functions to the~resonant points can be realized efficiently in two steps. Independence of the~frequency ratio on the~mass parameter $M$ implies the~conditions
\begin{equation}
      \nu_{\mathrm{U}}^{\mathrm{out}}(x;M,a) : \nu_{\mathrm{L}}^{\mathrm{out}}(x;M,a) = p^{\mathrm{out}}\,, \qquad
      \nu_{\mathrm{U}}^{\mathrm{in}}(x;M,a) : \nu_{\mathrm{L}}^{\mathrm{in}}(x;M,a) = p^{\mathrm{in}}
\end{equation}
giving relations for the~spin $a$ in terms of the~dimensionless radius $x$ and the~resonant frequency ratio $p$. They can be expressed in the~form $a^{\mathrm{out}}(x,p^{\mathrm{out}})$ and $a^{\mathrm{in}}(x,p^{\mathrm{in}})$, or in an~inverse form $x^{\mathrm{out}}(a,p^{\mathrm{out}})$ and $x^{\mathrm{in}}(a,p^{\mathrm{in}})$. At the~resonant radii the~conditions
\begin{equation}
        \nu^{\mathrm{out}}_{\mathrm{U}} = \nu^{\mathrm{out}}_{\mathrm{U}}(x;M,a)\,, \qquad    \nu^{\mathrm{in}}_{\mathrm{U}} = \nu^{\mathrm{in}}_{\mathrm{U}}(x;M,a)
\end{equation}
are satisfied along the~functions $M^{\mathrm{out}}_{p_{\mathrm{out}}}(a)$ and $M^{\mathrm{in}}_{p_{\mathrm{in}}}(a)$ which are obtained by using the~functions $a^{\mathrm{out}}(x,p^{\mathrm{out}})$ and $a^{\mathrm{in}}(x,p^{\mathrm{in}})$. The~parameters of the~neutron (quark) star are then given by the~condition
\begin{equation}
        M^{\mathrm{out}}_{p_{\mathrm{out}}}(a) = M^{\mathrm{in}}_{p_{\mathrm{in}}}(a)\,.    \label{RS}
\end{equation}
The~condition (\ref{RS}) determines $M$ and $a$ precisely, if the~resonant frequencies are determined precisely. If an~error occurs in determination of the~resonant frequencies, as naturally expected, our method gives related intervals of acceptable values of mass and spin parameter of the~neutron (quark) star.


\section{Resonant switch model applied to the~source 4U~1636$-$53}

We test the RS model in the~case of the~atoll 4U~1636$-$53 source that seems to be the~best possibility due to the~character of the~observational data demonstrating clearly existence of two ``resonant points'' where the~energy switch effect occurs. Using the~results of \cite{Tor:2009:ASTRA:ReversQPOs}, the~resonant frequencies determined by the~energy switch effect are given in the~outer resonant point with frequency ratio $\nu_{\mathrm{U}}/\nu_{\mathrm{L}}=3/2$ by the~frequency intervals
\begin{equation}
        \nu_{\mathrm{U}}^{\mathrm{out}} = \nu_{\mathrm{U0}}^{\mathrm{out}} \pm \Delta \nu^{\mathrm{out}} = (970 \pm 30) \,\mathrm{Hz}\,,\qquad
        \nu_{\mathrm{L}}^{\mathrm{out}} = \nu_{\mathrm{L0}}^{\mathrm{out}} \pm \Delta \nu^{\mathrm{out}} = (647 \pm 20) \,\mathrm{Hz}\,,  \label{freu}
\end{equation}
and at the~inner resonant point with frequency ratio $\nu_{\mathrm{U}}/\nu_{\mathrm{L}}=5/4$ there is
\begin{equation}
        \nu_{\mathrm{U}}^{\mathrm{in}} = \nu_{\mathrm{U0}}^{\mathrm{in}} \pm \Delta \nu^{\mathrm{in}} =  (1180 \pm 20) \,\mathrm{Hz}\,,\qquad
        \nu_{\mathrm{L}}^{\mathrm{in}} = \nu_{\mathrm{L0}}^{\mathrm{in}} \pm \Delta \nu^{\mathrm{in}} =   (944 \pm 16) \,\mathrm{Hz}\,. \label{frei}
\end{equation}
We consider the~standard specific models of the~twin oscillations based on the~orbital and epicyclic geodetical frequencies and in the framework of the RS model we determine range of allowed values of spin $a$ and mass $M$ of the~neutron star located at the~atoll source 4U~1636$-$53.

The~predicted ranges of the~neutron star parameters are strongly dependent on the~twin modes applied in the~RS~model. The mass and spin estimates of the RS model have to be confronted with restrictions on the neutron star mass and spin implied by the theoretical models of neutron (quark) stars and the observed rotational frequency of the neutron star at 4U~1636$-$53. We consider as acceptable upper values $a_{\mathrm{max}}\sim 0.4$, $M_{\mathrm{max}}\sim 2.5\, \mathrm{M}_{\odot}$ and demonstrate that for some of the~oscillatory modes used in the~RS~model the~predicted parameters of the~neutron star are unacceptable, being too high. Among acceptable RS~models the~most promising are those combining the~relativistic precession (where $\nu_{\mathrm{U}}=\nu_\mathrm{K}$, $\nu_{\mathrm{L}}=\nu_\mathrm{K} - \nu_r$) and the~total precession (where $\nu_{\mathrm{U}}=\nu_{\theta}$, $\nu_{\mathrm{L}}=\nu_{\theta} - \nu_r$) frequency relations or their modifications.
The switch of the oscillating modes is not given by the resonant phenomena
necessarily. The cause could be, e.g., given by the magnetic field, then
the Alfv\'{e}n wave model (\cite{Zhang04,Zhang06}) can be relevant.
\\\\
\textbf{Acknowledgements.} We thank grant GA\v{C}R~202/09/0772 and the project CZ.1.07/2.3.00/20.0071 ``Synergy'' supporting international collaboration of the Institute of Physics at SU Opava.

%

\end{document}